\documentclass[prd,a4paper,superscriptaddress,nofootinbib,10pt]{revtex4}
\usepackage{eurosym}
\usepackage{amsfonts}
\usepackage{graphicx}
\usepackage{amssymb}
\usepackage{amsmath}
\usepackage{lscape}
\usepackage{mathrsfs}
\usepackage{textcomp}
\usepackage{epsfig}
\usepackage{color}

\begin{document}

\title{Central tetrads and quantum spacetimes}
\author{Andrzej Borowiec}
\email{andrzej.borowiec@ift.uni.wroc.pl}
\affiliation{Institute of Theoretical Physics, University of Wroclaw, pl. M. Borna 9,
50-204 Wroclaw, Poland}
\author{Tajron Juri\'c}
\email{tjuric@irb.hr}
\author{Stjepan Meljanac}
\email{meljanac@irb.hr}
\affiliation{Rudjer Bo\v{s}kovi\'c Institute, Bijeni\v cka c.54, HR-10002 Zagreb, Croatia}
\author{Anna Pacho{\l }}
\email{apachol@unito.it}
\affiliation{Dipartimento di Matematica "Giuseppe Peano", Universita degli Studi di
Torino, Via Carlo Alberto, 10 - 10123 Torino, Italy}

\begin{abstract}
In this paper we perform a parallel analysis to the model proposed in \cite%
{majidbeggs}. By considering the central co-tetrad (instead of the central
metric) we investigate the modifications in the gravitational metrics coming
from the noncommutative spacetime of the $\kappa$-Minkowski type in four
dimensions. The differential calculus corresponding to a class of Jordanian $%
\kappa$-deformations provides metrics which lead either to cosmological
constant or spatial-curvature type solutions of non-vacuum Einstein
equations. Among vacuum solutions we find pp-wave type.
\end{abstract}

\maketitle

\section{Introduction}

The quantum gravity effects at the Planck scale might modify the structure
of spacetime leading to its noncommutativity \cite{DFR94,DFR95}. From
algebraic point of view in the quantum phase space, besides the non trivial
Heisenberg relations between coordinates and momenta, the coordinate
relations will be modified and one has to introduce noncommutativity of
coordinates themselves \cite{2,Zakrzewski}. Such modification of spacetime
might have influence on physical solutions, e.g. in gravitational and
cosmological effects \cite{Schupp, Schenkel, Mairi, BTZ}.
%In the past years, our knowledge on the universe has become more
%and more detailed thanks to new technologies which provide us with
%very precise astrophysical data.
The (noncommutative) modification of spacetime should be therefore included
in the theoretical predictions for (astrophysical) measurements \cite{link}.
Understandably any corrections to classical solutions would be of the order
of the Planck scale which makes them difficult to detect with today's
technology. However the theoretical models can suggest new directions to be
developed. %in experimental cosmology and astrophysics as well.
The finding of any falsifiable prediction to be tested in the real
(astrophysical) experiments and observations would be very important in the
experimental search for quantum gravity effects and high energy physics. The
considerations on deformation of gravitational solutions as well as on
cosmological effects coming from noncommutativity are also very timely due
to LIGO and PLANCK experiments.
%BICEP2's detection of B-mode polarization and
%PLANCK's results provide important observational input certain constraints
%on various cosmological models as well as which could be compared against
%the predictions coming from noncommutativity.

In noncommutative spacetimes approach it is assumed that effects of
noncommutativity should be visible in quantum gravity and would allow us to
model these in an effective description without full knowledge of quantum
gravity itself. One of the most known types of noncommutative spacetime is
when coordinates satisfy the Lie algebra type commutation relations. Such
deformation was inspired by the $\kappa$-deformed Poincar\'e algebra \cite{1}
as deformed symmetry for the $\kappa$-Minkowski spacetime \cite{2,
Zakrzewski}.

The investigations proposed in this paper focus on $\kappa $ type of
noncommutativity, where the $\kappa $-Minkowski commutation relations are as
follows:
\begin{equation}
\lbrack \hat{x}^{i},\hat{x}^{j}]=0,\quad \lbrack \hat{x}^{0},\hat{x}^{i}]=%
\frac{i}{\kappa }\hat{x}^{i},  \label{kappa1}
\end{equation}%
where $\kappa $ is the deformation parameter usually related to some quantum
gravity scale or Planck mass \cite{1,2}, for example certain bounds on this
scale were found in \cite{bgmp10, hajume}.

To study the modifications in the gravitational effects coming from this
deformed spacetime one needs to introduce the appropriate differential
calculus (compatible with such noncommutativity). There are few approaches
in constructing the deformed differential calculi. For example there are the
twisted approach \cite{Aschieri} and the bicovariant differential calculi
formulation based on quantum groups framework \cite{Woronowicz1}. Moreover
it has been shown that in the case of time-like $\kappa $-Minkowski
spacetime the four-dimensional bicovariant differential calculi compatible
with $\kappa $-Poincar\'e algebra does not exist, but one can construct a
five-dimensional one, which is bicovariant \cite{Sitarz, Gonera, Mercati}.
On the other hand considering light-like version of $\kappa $ -Minkowski
spacetime the differential calculus can be bicovariant and four-dimensional
\cite{hep-th/0307038, toward}. Differential calculi of classical dimension
(number of basis one-forms equal to number of coordinates) compatible with $\kappa
$ -Minkowski algebra (for time-, light- and spacelike deformations) were
classified in \cite{toward}. These differential calculi are bicovariant with
respect to other (larger) symmetries than $\kappa $-Poincar\'e algebra (except
for the light-like case). Alternative approaches to differential calculus on
$\kappa $-Minkowski space-time were also considered in \cite{Bu, KJ, 41,
EPJC, oeckl}.

Our aim in this paper, inspired by the papers \cite{majidbeggs, majid2014},
is to investigate the noncommutative (quantum) metrics coming from the
families of differential calculi introduced in \cite{41} (also included in
\cite{toward}).

In \cite{majidbeggs, majid2014} the authors have investigated the possible
noncommutative metrics $g$, which belong to the center of the $\kappa $%
-Minkowski algebra (\ref{kappa1}) for certain differential calculi, see also [27]. This
condition, necessary in the noncommutative Riemannian geometry to allow
contractions, defines set of equations for coefficients in the metric. In
the classical limit when $\kappa \rightarrow \infty $ the influence of the
noncommutativity remains in the form of the metric and leads to
modifications in the known solutions in GR. In this approach authors were
able to identify the Einstein tensor built from the central metric with that
of the perfect fluid for positive pressure, zero density, and for negative
pressure and positive density \cite{majidbeggs}. In a follow up paper \cite%
{majid2014} they showed that dark energy (cosmological constant) case can be
obtained from the algebraic constraint steaming from the central metric
approach.

Our aim in this paper is to consider the central tetrad fields $\omega^a$ in
the $\kappa$-Minkowski algebra (\ref{kappa1}) instead of the central metric.
This way the metric $g=\eta_{ab}\omega^a\otimes \omega^b$ has a Lorentzian
signature by definition provided that the flat metric $\eta_{ab}$ has the
same signature. In the central metric formalism one has to impose the
Lorentzian signature condition as an additional constraint.

Once we calculate the tetrads related with certain differential calculus
(compatible with the $\kappa$-Minkowski algebra) we can consider the
corresponding gravitational metric in the classical limit as induced from
noncommutativity. Classical limit is obtained by $\kappa \longrightarrow
\infty $ and then noncommutative objects (coordinates, differentials etc.)
will become commutative as follows:
\begin{equation}
\begin{split}
& \hat{x}^{k}\longrightarrow x^{k}\quad ,\quad \hat{x}^{0}\longrightarrow t
\\
& \hat{\xi}^{k}\longrightarrow \text{d}x^{k}\quad ,\quad \hat{\xi}%
^{0}\longrightarrow \text{d}t \\
& g=\hat{g}_{\mu \nu }\hat{\xi}^{\mu }\otimes \hat{\xi}^{\nu
}\longrightarrow g_{\mu \nu }\text{d}x^{\mu }\otimes \text{d}x^{\nu }
\end{split}%
\end{equation}
Of course, we impose that the metric derived from central tetrad in the
classical limit has to satisfy Einstein equations.

Then we focus on the non-vacuum Einstein equations in orthonormal tetrad form $G^{a b }=8\pi G T^{a b
} $ with $G^{a b} =R^{a b}-\frac{1}{2} R \eta^{a b}$ and the
energy momentum tensor $T^{a b}=(\rho, p, p, p)$ corresponding to the perfect isotropic and barotropic  fluid. In
cosmology, the equation of state of a perfect fluid is characterized by a
dimensionless number, the so-called barotropic factor $w$ equal to the ratio
of its pressure $p$ to the energy density $\rho $: $w=p/\rho $. For example
the most known cases are: $w=-1$ (cosmological constant or dark energy), $w=0$ (dust or dark matter),
$w=1/3$ (radiation) and $w=1$ (stiff matter). The value $w=-1/3$ corresponds to spatial curvature and separates two cases:
for $w\geqslant -1/3$ the strong energy condition $\rho+3p\geq 0$ is preserved, for $w<-1/3$ it is violated.
The last case characterizes accelerating universe while the former decelerating one.

The main result of this paper is that the effects of the noncommutativity
are encoded in the constraints coming from the central tetrad formalism
which induces a very special and generic classical solutions: universe
with a spatial curvature type of barotropic factor $w=-\frac{1}{3}$  and a
universe with dark energy (cosmological constant) with barotropic factor $%
w=-1$.
%The spatial curvature of the Friedmann-Lemaitre-Robertson-Walker (FLRW) metric is related with the so-called strong energy condition .

\section{$\protect\kappa$-Minkowski algebra and related quantum differential calculus%
}

\subsection{$\protect\kappa$-Minkowski algebra}

$\kappa$-Minkowski algebra $\hat{\mathcal{A}}=\mathbb{C}[\hat{x}^{\mu }]/%
\mathcal{I}$ with $\mu =0,..,3$ where $\mathcal{I}$ is a two-sided ideal
generated by the commutation relations
\begin{equation}  \label{kappa}
[\hat{x}^{i},\hat{x}^{j}]=0, \quad [\hat{x}^{0},\hat{x}^{i}]=\frac{i}{\kappa}%
\hat{x}^{i},
\end{equation}
where $\kappa$ is the deformation parameter usually related to some quantum
gravity scale or Planck mass. Eq. \eqref{kappa} represents the time-like
deformations of the usual Minkowski space. We can also look at more general
Lie algebraic deformations of Minkowski space
\begin{equation}  \label{kappalie}
[\hat{x}^{\mu},\hat{x}^{\nu}]=iC^{\mu\nu}{}_\lambda \hat{x}^{\lambda},
\end{equation}
where $\hat{x}^{\mu}=(\hat{x}^{0},\hat{x}^{i})$ and structure constants $%
C^{\mu\nu}{}_\lambda$ satisfy
\begin{equation}
C^{\mu\alpha}{}_\beta C^{\nu\lambda}{}_\alpha+C^{\nu\alpha}{}_\beta
C^{\lambda\mu}{}_\alpha +C^{\lambda\alpha}{}_\beta C^{\mu\lambda}{}_\alpha=0.
\end{equation}
\begin{equation}
{C^{\mu\nu}}_\lambda =- {C^{\nu\mu}}_\lambda.
\end{equation}
For $C^{\mu\nu}{}_\lambda=a^{\mu}\delta_{\lambda}^{\nu}-a^{\nu}\delta_{%
\lambda}^{\mu}$ we get
\begin{equation}  \label{kappaminkowski}
[\hat x^\mu, \hat x^\nu] = i(a^\mu\hat x^\nu - a^\nu\hat x^\mu).
\end{equation}
Generally, $a^\mu={\frac{1}{\kappa}} u^\mu$, where a fixed vector $u^\mu \in
\mathbb{R}^4$ belongs to the classical space (Minkowski space).
%, $a_\mu=\frac{%1}{\kappa}u_\mu$
For $u^{\mu}=(1,\vec{0})$ we get back to eq. \eqref{kappa} as a special
case. It turns out to be the most general case either since by linear change
of generators one can always transform relations \eqref{kappaminkowski} into %
\eqref{kappa} (cf. \cite{BP2014}). We should be also aware that these
relations and hence corresponding noncommutative algebras representing
quantum spacetimes are metric independent. Dependence of a metric may come
from covariance property under suitable quantum group action. For example,
taking the $\kappa-$Poincar\'e (quantum) group with metric of Lorentzian
signature $(-,+,+,+)$ one has to distinguish three cases, where $u^2=-1$ for
time-like deformations, $u^2=0$ for light-like deformations and $u^2=1$ for
space-like deformations.

\subsection{Differential calculus of classical dimension}

We denote the algebra of differential 1-forms as $d\hat{x}^{\mu }\in \Omega
^{1}( \hat{\mathcal{A}}) $ with $d:\hat{\mathcal{A}}\rightarrow \Omega ^{1}(
\hat{\mathcal{A}}) $. We define the basis 1-forms $d\hat{x}^{\mu }\equiv
\hat{\xi}^{\mu }$ in a usual way, where $d$ is the exterior derivative with
the property $d^{2}=0$ and satisfies Leibniz rule.

In \cite{toward} the construction of the most general algebra of
differential one-forms $\hat{\xi}^{\mu }$ compatible with $\kappa $%
-Minkowski algebra (\ref{kappa}) that is closed in differential forms (the
differential calculus is of classical dimension) is presented. The
commutators between one forms and coordinates are given by
\begin{equation}
\lbrack \hat{\xi}^{\mu },\hat{x}^{\nu }]=iK^{\mu }{}^{\nu}_{\alpha }\hat{\xi}%
^{\alpha },  \label{forme}
\end{equation}%
where $K^{\mu }{}^{\nu}_{\alpha }\in \mathbb{R}$ and after imposing
super-Jacobi identities and compatibility condition\footnote{%
for more details see \cite{toward}.} one gets two constraints on $K^{\mu
}{}^{\nu}_{\alpha }$
\begin{equation}
K^{\lambda \mu }{}_{\alpha }K^{\alpha \nu }{}_{\rho }-K^{\lambda \nu
}{}_{\alpha }K^{\alpha \mu }{}_{\rho }=C^{\mu \nu }{}_{\beta }K^{\lambda
\beta }{}_{\rho }.  \label{conditionK}
\end{equation}%
\begin{equation}
K^{\mu \nu }{}_{\alpha }-K^{\nu \mu }{}_{\alpha }=C^{\mu \nu }{}_{\alpha }.
\label{consist}
\end{equation}%
There are only four solutions (three of them are one parameter solutions) to
the above equations, which the authors of \cite{toward} denoted by $\mathcal{%
C}_{1}$, $\mathcal{C}_{2}$, $\mathcal{C}_{3}$ and $\mathcal{C}_{4}$ ($\mathcal{C}_{4}$ is valid
only for $a^{2}=0$). In this paper we will be interested in the original $%
\kappa $-Minkowski algebra, that is $a^{2}\neq 0$, and we are left with just
three special cases of $\mathcal{C}_{1}$, $\mathcal{C}_{2}$, $\mathcal{C}%
_{3} $. So, if we take $a^{2}\neq 0$ we get three families of algebras which
we denote by\footnote{%
Where $\mathcal{D}$ stands for $differential\ \ algebra$ and was already
used in \cite{41}.} $\mathcal{D}_{1}$, $\mathcal{D}_{2}$ and $\mathcal{D}%
_{3} $:
\begin{align}
& \mathcal{D}_{1}:~~~~[\hat{\xi}^{\mu },\hat{x}^{\nu }]=i\frac{s}{a^{2}}%
a^{\mu }a^{\nu }(a\hat{\xi})-ia^{\nu }\hat{\xi}^{\mu } \\
& \mathcal{D}_{2}:~~~~[\hat{\xi}^{\mu },\hat{x}^{\nu }]=i\frac{s}{a^{2}}%
a^{\mu }a^{\nu }(a\hat{\xi})-isa^{\nu }\hat{\xi}^{\mu }+i(1-s)a^{\mu }\hat{%
\xi}^{\nu } \\
& \mathcal{D}_{3}:~~~~[\hat{\xi}^{\mu },\hat{x}^{\nu }]=i\frac{s}{a^{2}}%
a^{\mu }a^{\nu }(a\hat{\xi})-i(1+s)\eta ^{\mu \nu }(a\hat{\xi})-ia^{\nu }%
\hat{\xi}^{\mu }
\end{align}%
For $a^{\mu }=(\frac{1}{\kappa },\vec{0})$ algebras $\mathcal{D}_{1}$ and $%
\mathcal{D}_{2}$ can be found in \cite{41}. For $s=1$ we see that algebras $%
\mathcal{D}_{1}^{s=1}$ and $\mathcal{D}_{2}^{s=1}$ coincide. This case was
in detail investigated in \cite{EPJC}. In \cite{oeckl} (see Corollary 5.1.)
the cases $\mathcal{D}_{1}^{s=0}$ and $\mathcal{D}_{2}^{s=0}$ were obtained
from a different construction.

In this paper we will focus on the differential algebra $\mathcal{D}_{1}$,
since the algebras $\mathcal{D}_{2,3}$ will only lead to degenerate central
metric. For $a^{\mu }=(\frac{1}{\kappa },\vec{0})$, $\mathcal{D}_{1}$ family
of algebra of differential forms has the following commutation relations
with $\kappa $-Minkowski coordinates (\ref{kappa}):
\begin{equation}
\left[ \hat{\xi}^{0},\hat{x}^{0}\right] =\frac{i}{\kappa }s\hat{\xi}%
^{0};\qquad \left[ \hat{\xi}^{k},\hat{x}^{0}\right] =-\frac{i}{\kappa }\hat{%
\xi}^{k};\qquad \left[ \hat{\xi}^{\mu },\hat{x}^{j}\right] =0
\label{family1}
\end{equation}%
where $s\in \mathbb{R}$ is a free parameter.

Let us consider a one-parameter family of Drinfeld Jordanian twist (cf. \cite%
{BP2009})
\begin{equation}  \label{twist}
\mathcal{F}=\exp \left\{ \ln Z\otimes \left( \frac{1}{s-1}%
L_{i}^{i}-L_{0}^{0}\right) \right\}
\end{equation}%
where $Z=[1+\frac{s-1}{\kappa }P_{0}]$ and $L_{\beta }^{\alpha }$ are
generators of the Lie algebra $\mathfrak{igl}$ of inhomogeneous general
linear transformations with the commutation relations $[L_{\nu }^{\mu
},L_{\lambda }^{\rho }]=\delta _{\nu }^{\rho }L_{\lambda }^{\mu }-\delta
_{\lambda }^{\mu }L_{\nu }^{\rho };\quad \lbrack L_{\nu }^{\mu },P_{\lambda
}]=-\delta _{\lambda }^{\mu }P_{\nu }$. These twists provide from one hand $%
\kappa -$Minkowski spacetime algebra (\ref{kappa}) and from the other a
family of differential calculus (\ref{family1}) in such a way that they are
%The family of differential calculus (\ref{family1}) is a
bicovariant with respect to the action of $U^{\mathcal{F}}(\mathfrak{igl)}$%
-Hopf algebra (for more details see \cite{toward}, \cite{EPJC}).

\section{Central co-tetrad and the corresponding deformed metric}

In this section we want to build up the quantum metric tensor from quantum
(noncommutative) co-tetrad $\hat{\omega}^{a}:$
\begin{equation}  \label{tetradmetric}
\hat{g}=\eta _{ab}\hat{\omega}^{a}\otimes \hat{\omega}^{b}=\hat{\omega}%
^{a}\otimes \hat{\omega}_{a}
\end{equation}%
where the flat classical metric $\eta _{ab}=diag\left( -,+,+,+\right) $ is
assumed to bear Lorentzian signature. In the classical limit co-tetrad
consists of four linearly independent one-forms $\omega^a=e^a_\mu dx^\mu$
which by construction are orthonormal with respect to the metric (\ref%
{tetradmetric}). The dual object composed of vector fields (named tetrad or
vierbein): $e_a=e_a^\mu\partial_\mu$ stands for famous E. Cartan \textit{rep%
\'{e}re mobile}. The matrices defining the tetrad $e_a^\mu$ and co-tetrad $%
e^a_\mu$ have to be mutually inverse each other. In fact, this approach
provides an effective link between flat and curved spacetime formalism and
is also useful in noncommutative setting (see e.g. \cite{Vitale} and
references therein). In addition the metric signature is controlled by the
signature of a flat metric.

In \cite{majidbeggs} the authors investigated the differential calculi
compatible with $\kappa $-Minkowski algebra of the classical dimension. In
our notation this type of differential calculus corresponds to the family $%
\mathcal{D}_{1}^{s=0}$. Later on \cite{majid2014} they also investigated
certain 2-dimensional differential calculi and they extended their
investigations to $\mathcal{D}_{1}$ and $\mathcal{D}_{2}$ families (which
was called $\alpha $ and $\beta $ family there). They showed that the $%
\mathcal{D}_{2}$ case leads to the degenerate central metrics, except for
the case: $\mathcal{D}_{1}^{s=1}\equiv \mathcal{D}_{2}^{s=1}$. In our
approach also the family $\mathcal{D}_{3}$ lead to degenerate metric as
well. Therefore in this section we look for other solutions in the $\mathcal{%
D}_{1}$ family.

We define the central co-tetrad $\hat{\omega}^{a}$ as a collection of four
linearly-independent one-forms that commute with all the noncommutative
coordinates $\hat{x}^{\mu }$ i.e.
\begin{equation}  \label{commtetrad}
\lbrack \hat{\omega}^{a},\hat{x}^{\mu }]=0.
\end{equation}%
and $\hat{\omega}^{a}$ can be written as
\begin{equation}
\hat{\omega}^{a}=e_{\mu }^{a}\hat{\xi}^{\mu },
\end{equation}%
where the components $e_{\mu }^{a}$ are functions of noncommutative
coordinates ($\in \hat{\mathcal{A}}$) that are yet to be determined.
Moreover, in the classical limit the matrix $e_{\mu }^{a}$ has to be
invertible (which provides an additional condition on functions $e_{\mu
}^{a} $).

In the algebra of $\kappa$-Minkowski coordinates the commutator (\ref%
{commtetrad}) is given by:
\begin{equation}
\left[ \hat{\omega}\left( \hat{x}^{\mu }\right) ,\hat{x}^{0}\right] =-\frac{i%
}{\kappa }\sum_{i}\hat{x}^{i}\frac{\partial }{\partial \hat{x}^{i}}\hat{%
\omega} \left( \hat{x}^{j},\hat{x}^{0}\right) ;\qquad \left[ \hat{\omega}%
\left( \hat{x}^{\mu }\right) ,\hat{x}^{k}\right] =\hat{x}^{k}\left( \omega
\left( \hat{x}^{i},\hat{x}_{0}+\frac{i}{\kappa }\right) -g\left( \hat{x}%
^{\mu }\right) \right)  \label{diff_eq}
\end{equation}

We start with a single central one-form $\hat{\omega}=e_{\mu }\xi ^{\mu }$
such that $[\hat{\omega},\hat{x}^{\alpha }]=0$.%
\begin{equation}
\lbrack \hat{\omega},\hat{x}^{0}]=[e_{\mu }\hat{\xi}^{\mu },\hat{x}%
^{0}]=e_{0}\left[ \hat{\xi}^{0},\hat{x}^{0}\right] +e_{k}\left[ \hat{\xi}%
^{k},\hat{x}^{0}\right] +[e_{0},\hat{x}^{0}]\hat{\xi}^{0}+[e_{k},\hat{x}^{0}]%
\hat{\xi}^{k}=0
\end{equation}%
\begin{equation}
\lbrack \hat{\omega},\hat{x}^{j}]=[e_{\mu }\hat{\xi}^{\mu },\hat{x}%
^{j}]=e_{0}\left[ \hat{\xi}^{0},\hat{x}^{j}\right] +e_{k}\left[ \hat{\xi}%
^{k},\hat{x}^{j}\right] +[e_{0},\hat{x}^{j}]\hat{\xi}^{0}+[e_{k},\hat{x}^{j}]%
\hat{\xi}^{k}=0
\end{equation}%
After using the relations (\ref{family1}) of differential calculus we get
the following conditions:%
%\bigskip \%$\left( e_{0}\frac{i}{\kappa }s+[e_{0},\hat{x}^{0}]\right) \hat{%
%\xi}^{0}+\left( [e_{k},\hat{x}^{0}]\pm \frac{i}{\kappa }e_{k}\right) \hat{\xi%
%}^{k}=0$
\begin{equation}
\lbrack e_{0},\hat{x}^{0}]=-\frac{i}{\kappa }se_{0}\quad ;\quad \lbrack
e_{k},\hat{x}^{0}]=\frac{i}{\kappa }e_{k}
\end{equation}%
and
\begin{equation}
\lbrack e_{0},\hat{x}^{j}]=0\quad ;\quad \lbrack e_{k},\hat{x}^{j}]=0
\end{equation}%
Therefore for the algebra $\mathcal{D}_{1}$ the requirement that the tetrad $%
\hat{\omega}$ is a central element in $\hat{\mathcal{A}}$ leads to the
following commutation relations
\begin{equation}
\begin{split}
& [e_{0},\hat{x}^{0}]=-\frac{i}{\kappa }se_{0}\quad ;\quad \lbrack e_{0},%
\hat{x}^{k}]=0 \\
& [e_{k},\hat{x}^{0}]=\frac{i}{\kappa }e_{k}\quad ;\quad \lbrack e_{k},\hat{x%
}^{i}]=0
\end{split}%
\end{equation}%
It turns out that all $e_{\mu }$ are only functions of $\hat{x}^{i}\,\ $(as
they commute with $\hat{x}^{j}$) and have to satisfy the following
differential equations (cf. \ref{diff_eq}):
\begin{equation}\label{25}
\hat{x}^{k}\frac{\partial }{\partial \hat{x}^{k}}e_{0} =se_{0}
\end{equation}
\begin{equation} \label{rt}
\hat{x}^{k}\frac{\partial }{\partial \hat{x}^{k}}e_{j} =-e_{j}
\end{equation}
The solutions of \eqref{25} and \eqref{rt} are (see also Appendix 1):
\begin{eqnarray}
e_{0} &=&\hat{r}^{s}E_{0}\left( \frac{\hat{x}^{k}}{\hat{r}}\right) \\
e_{j} &=&\hat{r}^{-1}E_{j}\left( \frac{\hat{x}^{k}}{\hat{r}}\right)
\end{eqnarray}%
So the central one-form reads as:
\begin{equation}
\hat{\omega}=\hat{r}^{s}E_{0}\left( \frac{\hat{x}^{k}}{\hat{r}}\right) \hat{%
\xi}^{0}+\hat{r}^{-1}E_{j}\left( \frac{\hat{x}^{k}}{\hat{r}}\right) \hat{\xi}%
^{j}
\end{equation}%
Similarly for the collection of four linearly independent one-forms
satisfying $[\hat{\omega}^{a},\hat{x}^{\alpha }]=0$ we get the following
solution:%
\begin{equation}  \label{tetradmetric1}
\hat{\omega}^{a}=\hat{r}^{s}E_{0}^{a}\left( \frac{\hat{x}^{k}}{\hat{r}}%
\right) \hat{\xi}^{0}+\hat{r}^{-1}E_{j}^{a}\left( \frac{\hat{x}^{k}}{\hat{r}}%
\right) \hat{\xi}^{j}
\end{equation}%
because
\begin{equation}
e_{0}^{a}=\hat{r}^{s}E_{0}^{a}\left( \frac{\hat{x}^{k}}{\hat{r}}\right)
\quad ;\quad e_{j}^{a}=\hat{r}^{-1}E_{j}^{a}\left( \frac{\hat{x}^{k}}{\hat{r}%
}\right)
\end{equation}%
In result we have 16 arbitrary functions $E_{\mu }^{a}\left( \frac{\hat{x}%
^{k}}{\hat{r}}\right) $ of variables $\frac{\hat{x}^{k}}{\hat{r}}$ and $\hat{%
r}=\sqrt{\hat{x}^{k}\hat{x}_{k}}$ (note that in spherical coordinates such
functions will only depend on the angles).

Now the metric can be built up from the central tetrads (\ref{tetradmetric})
as follows:
\begin{eqnarray}
\hat{g} &=&\left( \hat{r}^{s}E_{0}^{a}\left( \frac{\hat{x}^{k}}{\hat{r}}%
\right) \hat{\xi}^{0}+\hat{r}^{-1}E_{i}^{a}\left( \frac{\hat{x}^{k}}{\hat{r}}%
\right) \hat{\xi}^{i}\right) \otimes \left( \hat{r}^{s}E_{a0}\left( \frac{%
\hat{x}^{k}}{\hat{r}}\right) \hat{\xi}^{0}+\hat{r}^{-1}E_{aj}\left( \frac{%
\hat{x}^{k}}{\hat{r}}\right) \hat{\xi}^{j}\right) = \\
&=&\hat{r}^{2s}E_{0}^{a}E_{a0}\hat{\xi}^{0}\otimes \hat{\xi}^{0}+\hat{r}%
^{s-1}E_{a0}E_{j}^{a}\hat{\xi}^{0}\otimes \hat{\xi}^{j}+\hat{r}%
^{s-1}E_{i}^{a}E_{a0}\hat{\xi}^{i}\otimes \hat{\xi}^{0}+\hat{r}%
^{-2}E_{i}^{a}E_{aj}\hat{\xi}^{i}\otimes \hat{\xi}^{j}
\end{eqnarray}
One can show that this type of the metric belongs to the center of the
algebra $\hat{\mathcal{A}}$ of $\kappa $-Minkowski type (\ref{kappa}) as
well and has vanishing commutators:
\begin{equation}
\left[ \hat{g},\hat{x}^{\mu }\right] =0  \label{metric_cent}
\end{equation}%
i.e. it falls in the framework introduced in \cite{majidbeggs}.

\subsection{Classical limit}

In the classical limit $\kappa \longrightarrow \infty $ when the
noncommutative objects (coordinates, differentials etc.) become commutative,
the functions $E_{\mu }^{a}$ will become the arbitrary functions of $\frac{{x%
}^{k}}{{r}}$ and $r=\sqrt{x^{k}x^{k}}$ commutative coordinates.

In this case the metric for algebra $\mathcal{D}_{1}$ (\ref{family1}) in the
classical limit reads as:
\begin{equation}
g=g_{\mu \nu }dx^{\mu }\otimes dx^{\nu }=\frac{1}{r^{2}}E_{i}^{a}E_{aj}{dx}%
^{i}\otimes {dx}^{j}+r^{\left( s-1\right) }E_{i}^{a}E_{a0}\left( {dx}%
^{i}\otimes {dt}+{dt}\otimes {dx}^{i}\right) +{r}^{2s}E_{0}^{a}E_{a0}{dt}%
\otimes {dt}
\end{equation}%
One can see that in the above metric the functions $E_{\mu }^{a}$ do not
depend on time therefore such metrics could describe only stationary solutions.

%, e.g. excluding cosmological models.

When we introduce the spherical coordinates
\begin{equation*}
\left( t,r,\theta ,\phi \right):\quad x=r\sin \theta \cos \phi, \quad
y=r\sin \theta \sin \phi, \quad z=r\cos \theta
\end{equation*}
the functions $E_{\mu }^{a}\left(\theta,\phi \right) $ are arbitrary
functions of the angles $\left( \theta,\phi\right) $ only. Therefore, in
what follows, we shall use spherical coordinate system to write down the
metric in the form: \footnote{%
In fact, the coefficient functions $E^a_\mu (\theta, \phi)$ appearing in
formulas (35) and below are in general not identical. They can be express
each other as linear combinations with trigonometric functions of $(\theta,
\phi)$.}

\begin{center}
$g=\left(
\begin{array}{cccc}
r^{2s}E_{0}^{a}\left( \theta,\phi \right) E_{a0}\left( \theta,\phi \right) &
r^{\left( s-1\right) }E_{1}^{a}\left(\theta, \phi \right) E_{a0}\left(
\theta,\phi \right) & r^{s}E_{2}^{a}\left( \theta,\phi \right) E_{a0}\left(
\theta,\phi \right) & r^{s}E_{3}^{a}\left( \theta,\phi \right) E_{a0}\left(
\theta,\phi \right) \\
r^{\left( s-1\right) }E_{1}^{a}\left( \theta,\phi \right) E_{a0}\left(
\theta,\phi \right) & \frac{E_{1}^{a}\left( \theta,\phi \right) E_{a1}\left(
\theta,\phi \right) }{r^{2}} & \frac{E_{1}^{a}\left( \theta,\phi \right)
E_{a2}\left( \theta,\phi \right) }{r} & \frac{E_{1}^{a}\left( \theta,\phi
\right) E_{a3}\left( \theta,\phi \right) }{r} \\
r^{s}E_{2}^{a}\left( \theta,\phi \right) E_{a0}\left( \theta,\phi \right) &
\frac{E_{1}^{a}\left(\theta,\phi \right) E_{a2}\left( \theta,\phi \right) }{r%
} & E_{2}^{a}\left( \theta,\phi \right) E_{a2}\left( \theta,\phi \right) &
E_{2}^{a}\left( \theta,\phi \right) E_{a3}\left( \theta,\phi \right) \\
r^{s}E_{3}^{a}\left( \theta,\phi \right) E_{a0}\left( \theta,\phi \right) &
\frac{E_{1}^{a}\left( \theta,\phi \right) E_{a3}\left(\theta,\phi \right) }{r%
} & E_{2}^{a}\left( \theta,\phi \right) E_{a3}\left( \theta,\phi \right) &
E_{3}^{a}\left( \theta,\phi \right) E_{a3}\left( \theta,\phi \right)%
\end{array}%
\right) $
\end{center}

We are looking for non degenerate solutions so from the condition on
non-zero determinant $\det \left( g\right)\neq 0$ we get additional
constraints on the functions $E_{\mu }^{a}$.

We can simplify the notation  by introducing
\begin{eqnarray}
r^{-2}E_{i}^{a}\left( \theta,\phi \right) E_{aj}\left( \theta,\phi \right)
&=&r^{-2}a_{ij}\left( \theta,\phi \right) \\
r^{s-1}E_{0}^{a}\left( \theta,\phi \right) E_{aj}\left( \theta,\phi \right)
&=&r^{s-1}b_{j}\left( \theta,\phi \right) \\
r^{2s}E_{0}^{a}\left( \theta,\phi \right) E_{a0}\left( \theta,\phi \right)
&=&r^{2s}c\left( \theta,\phi \right)
\end{eqnarray}

The Einstein equations are written in the form ($G$  being the universal Newtonian constant of gravity):
\begin{equation}  \label{ee}
R_{\mu\nu}-\frac{1}{2}g_{\mu\nu}R=8\pi G\,T_{\mu\nu}
\end{equation}
Given a specified distribution of matter and energy $T_{\mu\nu}$, the
equations (\ref{ee}) are understood to be equations for the metric tensor $%
g_{\mu\nu}$, as both the Ricci tensor $R_{\mu\nu}$ and scalar curvature $R$
depend on the metric in a complicated nonlinear manner. One can write the
Einstein equations in a more compact form by introducing the Einstein tensor
$G_{\mu \nu }=R_{\mu \nu }-\frac{1}{2}g_{\mu \nu }R$. Since in the classical
limit our metric has to satisfy the Einstein equations, this leads us to
some coupled differential equations for tetrad functions $E_{\mu }^{a}$
which turn out to be non trivial to solve. In order to find some solutions,
we consider a special case with the 'constant coefficients', i.e. such that
the metric depends only on $r$ coordinate and:
\begin{equation}
E_{i}^{a}E_{aj}=a_{ij}=const\quad ;\quad E_{a0}E_{j}^{a}=b_{j}=const\quad
;\quad E_{0}^{a}E_{a0}=c=const
\end{equation}
The metric looks as follows:
\begin{equation}  \label{qmetric}
g=\left(
\begin{array}{cccc}
r^{2s}c & r^{\left( s-1\right) }b_{1} & r^{s}b_{2} & r^{s}b_{3} \\
r^{\left( s-1\right) }b_{1} & \frac{a_{11}}{r^{2}} & \frac{a_{12}}{r} &
\frac{a_{13}}{r} \\
r^{s}b_{2} & \frac{a_{12}}{r} & a_{22} & a_{23} \\
r^{s}b_{3} & \frac{a_{13}}{r} & a_{23} & a_{33}%
\end{array}%
\right)
\end{equation}
and the determinant: {\footnotesize {$r^{2s-2}\{b_{1}^{2}a_{23}^{2}-
b_{1}^{2}a_{22}a_{33}+b_{2}^{2}a_{13}^{2}-b_{2}^{2}a_{11}a_{33}+b_{3}^{2}
a_{12}^{2}-b_{3}^{2}a_{11}a_{22}-ca_{11}a_{23}^{2}-ca_{13}^{2}a_{22}-ca_{12}^{2}a_{33}+ca_{11}a_{22}a_{33}+2ca_{12}a_{13}a_{23}+2b_{1}b_{2}a_{12}a_{33}-2b_{1}b_{2}a_{13}a_{23}-2b_{1}b_{3}a_{12}a_{23}+2b_{1}b_{3}a_{13}a_{22}+\allowbreak 2b_{2}b_{3}a_{11}a_{23}-2b_{2}b_{3}a_{12}a_{13}\}\neq 0
$}} is assumed to be non-zero.

In the following we will look for the solutions of non-vacuum $G^{\mu \nu
}=8\pi G\,T^{\mu \nu }$ and vacuum $G^{\mu \nu }=0$ Einstein equations.

\subsubsection{Non vacuum solutions}

In the tetrad formalism we can work with Einstein equations in the Lorentzian
frame: $G^{ab}=8\pi G\,T^{ab}$. The passage to the coordinate frame is determined by
standard formulae: $g_{\mu \nu }=\eta _{ab}e_{\mu }^{a}e_{\nu }^{b}$ with $%
\eta _{ab}=diag\left( -,+,+,+\right) $ and $G_{\nu }^{\mu }=e_{a}^{\mu
}e_{\nu }^{b}G_{b}^{a}$, etc.

On the left hand side of the non-vacuum Einstein equations we assume the
energy-momentum of perfect and isotropic fluid, i.e. $T^{ab}=(\rho
+p)u^{a}u^{b}+p\eta ^{ab}=diag(\rho ,p,p,p)$ where $\rho $ is energy
density, $p$ is the pressure of the fluid and the vector $u^{a}$ represents
its four-velocity. Assuming further that the fluid is co-moving with respect
to the tetrad, i.e. $u^{a}=(1,0,0,0)$ we can diagonalize the energy-momentum
tensor $T_{b}^{a}=diag(-\rho ,p,p,p)$. Therefore to look for solutions with
perfect fluid one should firstly diagonalize the Einstein tensor $%
G_{\mu}^{\nu}.$ \footnote{%
The %\textquotedblright world\textquotedblright\
Lorentz indices $a,b.\ldots $ are raised and lowered by means of the flat
metric and its inverse.} For this purpose we calculate the characteristic
polynomial: $\det \left[ G_{\mu}^{\nu}-\lambda \delta _{\mu}^{\nu}\right]
\Leftrightarrow\det \left[ G_{\mu\nu}-\lambda g _{\mu\nu}\right]=0$ which
will give us the diagonal form of the $G_{b}^{a}$ with the roots of this
equation on the diagonal. Having the multiplicity $1$ of one solution $%
\tilde{\lambda}$ and multiplicity $3$ of another $\lambda $ will allow us to
write down the equation with the (diagonal) momentum energy tensor for the
perfect fluid as:
\begin{equation}
G_{b}^{a}=\left(
\begin{array}{cccc}
\tilde{\lambda} &  &  &  \\
& \lambda &  &  \\
&  & \lambda &  \\
&  &  & \lambda%
\end{array}%
\right) =8\pi G\,\left(
\begin{array}{cccc}
-\rho &  &  &  \\
& p &  &  \\
&  & p &  \\
&  &  & p%
\end{array}%
\right)
\end{equation}

The equation of state of barotropic fluid is characterized by a
dimensionless number $w$ equal to the ratio of its pressure $p$ to the
energy density $\rho $: $w=p/\rho =-\lambda /\tilde{\lambda}$. For example
the most known from cosmology cases are:

- cosmological constant (dark energy) which corresponds to $w=-1$,

- dust matter (dark or/and ordinary baryonic matter) ($w=0$),

- radiation ($w=1/3$).

\noindent It turns out that spatial curvature of FLRW metric can be also
described by the barotropic factor $w=-{\frac{1}{3}}$ satisfying strong
energy condition $\rho+3p\geq 0$.
%In our case we firstly check the non vacuum solutions with isotropic metric $%
%\ref{qmetric}$.

Therefore after the diagonalization of the Einstein tensor we can see which
kind of the equation of state can be derived from the quantum metric (\ref%
{qmetric}). In the case under consideration one obtains only two possible
solution of the barotropic type (there is no other solutions).

\begin{enumerate}
\item Quantum Universe with spatial curvature type barotropic factor $w=-1/3$%
;

$\left( G_{b}^{a}\right) _{_{I}}=\left(
\begin{array}{cccc}
\lambda ^{I} & 0 & 0 & 0 \\
0 & \frac{1}{3}\lambda ^{I} & 0 & 0 \\
0 & 0 & \frac{1}{3}\lambda ^{I} & 0 \\
0 & 0 & 0 & \frac{1}{3}\lambda ^{I}%
\end{array}%
\right) =8\pi G\,\left(
\begin{array}{cccc}
-\rho & 0 & 0 & 0 \\
0 & p & 0 & 0 \\
0 & 0 & p & 0 \\
0 & 0 & 0 & p%
\end{array}%
\right) $

$\left( G^{ab}\right) _{_{I}}=\left(
\begin{array}{cccc}
-\lambda ^{I} & 0 & 0 & 0 \\
0 & \frac{1}{3}\lambda ^{I} & 0 & 0 \\
0 & 0 & \frac{1}{3}\lambda ^{I} & 0 \\
0 & 0 & 0 & \frac{1}{3}\lambda ^{I}%
\end{array}%
\right) $

where $\lambda ^{I}=$ $-\frac{{a}_{{22}}s^{2}}{4\left( {a}_{{12}}^{2}-{a}_{{%
11}}{a}_{{22}}\right) }.$This eigenvalue is given for simplified metric with
the choice $c=0;b_{1}=0;b_{2}=0$, which does not change the barotropic
factor. The full expression for $\lambda ^{I}$ corresponding exactly to (\ref%
{qmetric}) can be found in Appendix 2. From "polynomial constraints" (see
Appendix 2) it follows that $a_{23}=\sqrt{a_{22}a_{33}}$ . The determinant
has to be non-zero $\det \left( g_{I_{simpl}}\right)
=r^{2s-2}b_{3}^{2}\left( a_{12}^{2}-a_{11}a_{22}\right) \neq 0$ and $\det
\left( g_{I_{simpl}}\right) <0$ iff $a_{12}^{2}<a_{11}a_{22}$. The
corresponding metric is then:

\begin{equation}
g_{I_{simpl}}=\left(
\begin{array}{cccc}
0 & 0 & 0 & r^{s}b_{3} \\
0 & \frac{a_{11}}{r^{2}} & \frac{a_{12}}{r} & \frac{a_{13}}{r} \\
0 & \frac{a_{12}}{r} & a_{22} & \sqrt{a_{22}a_{33}} \\
r^{s}b_{3} & \frac{a_{13}}{r} & \sqrt{a_{22}a_{33}} & a_{33}%
\end{array}%
\right)
\end{equation}

\item Universe with dark energy (cosmological constant) with $w=-1$;

$\left( G_{b}^{a}\right) _{II}=\left(
\begin{array}{cccc}
\lambda ^{II} & 0 & 0 & 0 \\
0 & \lambda ^{II} & 0 & 0 \\
0 & 0 & \lambda ^{II} & 0 \\
0 & 0 & 0 & \lambda ^{II}%
\end{array}%
\right) =8\pi G\,\left(
\begin{array}{cccc}
-\rho  & 0 & 0 & 0 \\
0 & p & 0 & 0 \\
0 & 0 & p & 0 \\
0 & 0 & 0 & p%
\end{array}%
\right) $

$\left( G^{ab}\right) _{II}=\left(
\begin{array}{cccc}
-\lambda ^{II} & 0 & 0 & 0 \\
0 & \lambda ^{II} & 0 & 0 \\
0 & 0 & \lambda ^{II} & 0 \\
0 & 0 & 0 & \lambda ^{II}%
\end{array}%
\right) $

where $\lambda ^{II}=\frac{s^{2}}{2{a}_{{11}}}$ (is the simplified version
for the choice of the coefficients $a_{12}=0;a_{23}=0;a_{13}=0;$ $%
b_{1}=0=b_{3}$). Note that $s\neq 0.$ The expression for $\lambda _{2}$
depending on all constant coefficients is given in the Appendix 2 as well.
The metric itself looks as follows:
\begin{equation}
g_{II_{simpl}}=\left(
\begin{array}{cccc}
r^{2s}\frac{{b}_{{2}}^{2}}{2{a}_{22}} & 0 & r^{s}b_{2} & 0 \\
0 & \frac{a_{11}}{r^{2}} & 0 & 0 \\
r^{s}b_{2} & 0 & a_{22} & 0 \\
0 & 0 & 0 & a_{33}%
\end{array}%
\right)
\end{equation}%
The determinant of the metric: $\det \left( g_{II_{simpl}}\right) =-\frac{1}{%
2}b_{2}^{2}a_{33}a_{11}\neq 0$ and $\det g_{II_{simpl}}$ can be chosen to be
negative for $a_{33}a_{11}>0$.
\end{enumerate}

However, choosing $s=0$ in this case, i.e. $\left(
\begin{array}{cccc}
\frac{{b}_{{2}}^{2}}{2{a}_{22}} & 0 & b_{2} & 0 \\
0 & \frac{a_{11}}{r^{2}} & 0 & 0 \\
b_{2} & 0 & a_{22} & 0 \\
0 & 0 & 0 & a_{33}%
\end{array}%
\right) $ we get an example of vacuum (but flat) solution of Einstein
equations, for which $G_{\mu \nu }=0$ (note that $G^{\mu \nu }=0\Rightarrow
R^{\mu \nu }=0$). Below we present yet other interesting and not flat vacuum
solutions, both for $s\neq 0$ and $s=0$ cases.

\subsubsection{Vacuum solutions}

\begin{enumerate}
\item Vacuum solution of the pp-wave type.

One of the examples of the vacuum solution can be provided for the special
choice of non-zero, constant coefficients $b_{1},a_{22},a_{33}$ and
restoring the dependence on the angles in the coefficient $a_{11}\left( \phi
,\theta \right) $ i.e. choosing the metric as:
\begin{equation}
g_{1vac}=\left(
\begin{array}{cccc}
0 & r^{\left( s-1\right) }b_{1} & 0 & 0 \\
r^{\left( s-1\right) }b_{1} & \frac{a_{11}\left( \phi ,\theta \right) }{r^{2}%
} & 0 & 0 \\
0 & 0 & a_{22} & 0 \\
0 & 0 & 0 & a_{33}%
\end{array}%
\right)  \label{vac1}
\end{equation}%
Then the corresponding Einstein tensor vanishes for any choice of the
parameter $s\neq 0$ but under the following additional conditions on the
derivatives of function $a_{11}\left( \phi ,\theta \right) $ as follows: $%
\frac{\partial a_{11}}{\partial \phi }\neq 0;\frac{\partial a_{11}}{\partial
\theta }\neq 0;\frac{\partial ^{2}a_{11}}{\partial \phi \partial \theta }%
\neq 0$ and $\frac{\partial ^{2}a_{11}}{\partial \phi ^{2}}=0=\frac{\partial
^{2}a_{11}}{\partial \theta ^{2}}$ [which amounts to $a_{11}\left( \phi
,\theta \right) =\xi _{1}+\xi _{2}\theta +\xi _{3}\phi +\xi _{4}\theta \phi $%
, $\xi _{i}$ are constants]. The Riemann tensor however does not vanish: $%
R_{t\theta \phi r}=\frac{r^{-\left( s+1\right) }}{2b_{1}}\frac{\partial
^{2}a_{11}}{\partial \phi \partial \theta }=R_{t\phi \theta r};R_{\theta
r\phi r}=-\frac{1}{2a_{22}r^{2}}\frac{\partial ^{2}a_{11}}{\partial \phi
\partial \theta };R_{\phi r\theta r}=-\frac{1}{2a_{33}r^{2}}\frac{\partial
^{2}a_{11}}{\partial \phi \partial \theta }$. Therefore (\ref{vac1})
constitutes vacuum but non-flat solution.

\item Another example of vacuum solution, also of pp-wave type can be
provided (for $s=0$), by:
%Again we do the special choice of non-zero, constant coefficients $b_{1},a_{22},a_{33}$ with restored dependence on the angles in
%the coefficient $c\left( \phi ,\theta \right) $ :
\begin{equation}
g_{2vac}=\left(
\begin{array}{cccc}
c\left( \phi ,\theta \right)  & r^{\left( -1\right) }b_{1} & 0 & 0 \\
r^{\left( -1\right) }b_{1} & 0 & 0 & 0 \\
0 & 0 & a_{22} & 0 \\
0 & 0 & 0 & a_{33}%
\end{array}%
\right)   \label{vac2}
\end{equation}%
The corresponding Einstein tensor vanishes by imposing additional conditions
(analogous to the above ones) on the derivatives of function $c\left( \phi
,\theta \right) $ as follows: $\frac{\partial c}{\partial \phi }\neq 0;\frac{%
\partial c}{\partial \theta }\neq 0;\frac{\partial ^{2}c}{\partial \phi
\partial \theta }\neq 0$ and$\frac{\partial ^{2}c}{\partial \phi ^{2}}=0=%
\frac{\partial ^{2}c}{\partial \theta ^{2}}$ [which amounts to $c\left(
\phi ,\theta \right) =\zeta _{1}+\zeta _{2}\theta +\zeta _{3}\phi +\zeta
_{4}\theta \phi ,\zeta _{i}$ are constants]. However the Riemann tensor does
not vanish ($R_{r\theta \phi t}=\frac{r}{2b_{1}}\frac{\partial ^{2}c}{%
\partial \phi \partial \theta }=R_{r\phi \theta t};R_{\theta t\phi t}=-\frac{%
1}{2a_{22}}\frac{\partial ^{2}c}{\partial \phi \partial \theta };R_{\phi
t\theta t}=-\frac{1}{2a_{33}}\frac{\partial ^{2}c}{\partial \phi \partial
\theta }$). Therefore (\ref{vac2}) is as well vacuum non-flat (Ricci-flat)
solution.

It is rather known that the so called pp-waves (plane-fronted waves with
parallel rays) are exact wave like solutions of Einstein equations which
represent gravitational radiation propagating with the speed of light in a
direction determined by light-like Killing vector field. They are analogous
to source-free photons in Maxwell electrodynamics. They also appear in many
places of Theoretical Physics e.g. as string or D-brane background, and
supersymmetry, etc..

\end{enumerate}

Note that the non-vacuum and vacuum solutions for $s=0$ are interesting from
the quantum symmetry point of view, since in this case the Jordanian twist (%
\ref{twist}) reduces to
\begin{equation}
\mathcal{F}=\exp \left\{ \ln \left( 1-\frac{1}{\kappa }P_{0}\right) \otimes
D\right\}
\end{equation}%
and corresponds to Poincar\'e-Weyl symmetry (one generator extension of the
Poincar\'e algebra, with adding dilatation generator $D$) as a minimal
quantum group providing symmetry algebra \cite{BP2009}.

\section{Final remarks}

We have investigated some of the properties of quantum spaces using the
central co-tetrad formalism. We used a sort of a toy model for quantum
gravity effects, which should be encoded in the noncommutative $\kappa $%
-Minkowski algebra \eqref{kappa} and by analyzing a certain bicovariant
differential calculus (compatible with such noncommutative structure) we
found equations for the components of the noncommutative central co-tetrads,
which we solve in general \eqref{tetradmetric1}. We show that this formalism
gives rise to the same quantum central metric as proposed in \cite%
{majidbeggs}, but here the benefit is that the Lorentzian signature is built
in from the very beginning.

Analyzing the classical limit of our quantum metric, and by imposing the
validity of Einstein equations in that limit, we found (under further
assumptions which simplify the calculations) new vacuum and non-vacuum
solutions. Namely, the solutions of Einstein equations for our simplified
cases (metric with constant coefficients and only $r$-dependence) contribute
to the description of the (quantum) Universe with the cosmological constant
and the spacial curvature. Also, a vacuum solution corresponding to pp-wave
spacetime is obtained.

The idea is to investigate more complicated and more general solution of %
\eqref{tetradmetric1}. One can see that in the quantum metric %
\eqref{tetradmetric1}, the functions $E_{\mu }^{a}$ do not depend on time
therefore such metrics could describe only static solutions for $\mathcal{D}%
_{1}$. Maybe for more general differential calculi $\mathcal{C}_{1}...,%
\mathcal{C}_{4}$ (classified in \cite{toward}) one can recover some new
interesting cosmological and maybe even black hole solutions.

\section*{Appendix 1}

We want to find the general solution for the following differential equation
\begin{equation}
\hat{x}^{i}\frac{\partial }{\partial \hat{x}^{i}}f=\gamma f  \label{master}
\end{equation}%
We denote the dilatation operator $\hat{D}=\hat{x}^{i}\frac{\partial }{%
\partial \hat{x}^{i}}$. For any dimension we can define the spherical
coordinates and write the radial vector and nabla operators as
\begin{equation}
\vec{\hat{r}}=\hat{x}_{k}\vec{e}_{k}=\hat{r}\ \vec{\hat{r}}_{0}\quad \vec{%
\nabla}=\vec{e}_{k}\frac{\partial }{\partial \hat{x}_{k}}=\vec{\hat{r}}_{0}%
\frac{\partial }{\partial \hat{r}}+\text{terms in directions perpendicular to%
}\ \vec{\hat{r}}_{0}
\end{equation}%
where $\vec{e}_{k}$ are constant orthonormal vectors of basis in Cartesian
coordinate system and $\vec{\hat{r}}_{0}$ is radial unit vector. Now, for
the dilatation operator we have
\begin{equation}
\hat{D}=\hat{x}^{i}\frac{\partial }{\partial \hat{x}^{i}}=\vec{\hat{r}}\cdot
\vec{\nabla}=\hat{r}\frac{\partial }{\partial \hat{r}}
\end{equation}%
so, differential equation \eqref{master} can be solved by direct integration
which gives the following solution
\begin{equation}
f=\text{const.}\ \hat{r}^{\gamma }
\end{equation}%
Notice that $\hat{D}\frac{x_{k}}{\hat{r}}=0$, and so any arbitrary function
of $\frac{x_{k}}{\hat{r}}$ satisfies $\hat{D}F(\frac{x_{k}}{\hat{r}})=0$ the
most general \textquotedblleft const.\textquotedblright\ is $F(\frac{x_{k}}{%
\hat{r}})$ which gives the most general solution for \eqref{master} is

\begin{equation}
f=\hat{r}^{\gamma }F\left( \frac{x_{k}}{\hat{r}}\right)
\end{equation}%
This enables us to solve differential equations.

\section*{Appendix 2: Full solution for the metric (\protect\ref{qmetric})}

In section III.A.1 we focused on non vacuum solutions with isotropic metric (%
\ref{qmetric}) , i.e. for $a_{ij}=const,b_{i}=const$ and $c=const)$ and only
with dependence on $r:$
\begin{equation}
g=\left(
\begin{array}{cccc}
r^{2s}c & r^{\left( s-1\right) }b_{1} & r^{s}b_{2} & r^{s}b_{3} \\
r^{\left( s-1\right) }b_{1} & \frac{a_{11}}{r^{2}} & \frac{a_{12}}{r} &
\frac{a_{13}}{r} \\
r^{s}b_{2} & \frac{a_{12}}{r} & a_{22} & a_{23} \\
r^{s}b_{3} & \frac{a_{13}}{r} & a_{23} & a_{33}%
\end{array}%
\right)
\end{equation}

After diagonalization of the Einstein tensor $G_{\nu }^{\mu }$ wrt this
metric we obtain only two possible solutions of the perfect fluid type
(there is no other solutions). The characteristic equation is of the form : $%
\det \left[ G_{\nu }^{\mu }-\lambda \delta _{\nu }^{\mu }\right] =\alpha
\cdot \beta \cdot \gamma =0$ where:\newline

$\alpha =\left( \left( -{a}_{{33}}{b}_{{2}}^{2}+2{a}_{{23}}{b}_{{2}}{b}_{{3}%
}-{a}_{{22}}{b}_{{3}}^{2}\right) s^{2}+{x\lambda }_{1}\right) ^{2}$\newline
$\beta =\left( \left( -3{a}_{{33}}{b}_{{2}}^{2}+6{a}_{{23}}{b}_{{2}}{b}_{{3}%
}-3{a}_{{22}}{b}_{{3}}^{2}-4{a}_{{23}}^{2}c+4{a}_{{22}}{a}_{{33}}c\right)
s^{2}+{x\lambda }_{2}\right) $\newline
$\gamma =\left( \left( -{a}_{{33}}{b}_{{2}}^{2}+2{a}_{{23}}{b}_{{2}}{b}_{{3}%
}-{a}_{{22}}{b}_{{3}}^{2}-4{a}_{{23}}^{2}c+4{a}_{22}{a}_{{33}}c\right)
s^{2}+x\lambda _{3}\right) $\newline

and $x=-4\left( {a}_{{23}}^{2}-{a}_{{22}}{a}_{{33}}\right) b_{{1}%
}^{2}-4\left( {a}_{{12}}^{2}-{a}_{{11}}{a}_{{22}}\right) b_{{3}}^{2}-4\left(
{a}_{{13}}^{2}-{a}_{{11}}{a}_{{33}}\right) b_{{2}}^{2}$

$+8a_{{13}}a_{{23}}b_{{1}}b_{{2}}-8a_{{12}}a_{{33}}b_{{1}}b_{{2}}-8a_{{13}%
}a_{{22}}b_{{1}}b_{{3}}+8a_{{12}}a_{{23}}b_{{1}}b_{{3}}$

$+8a_{{12}}a_{{13}}b_{{2}}b_{{3}}-8a_{{11}}a_{{23}}b_{{2}}b_{{3}}+4\left(
a_{13}^{2}a_{22}-2a_{12}a_{13}a_{23}+a_{11}a_{23}^{2}+a_{12}^{2}a_{33}-a_{11}a_{22}a_{33}\right) c
$\newline

I. To find the solution of multiplicity 3 and 1 (corresponding to subcase 1. in
Sec.III.A.1 ) we notice that for:  $-4{a}_{{23}}^{2}c+4{a}_{{22}}{a}_{{33}%
}c=0$ we have the following:

$\alpha =\left( \left( -{a}_{{33}}{b}_{{2}}^{2}+2{a}_{{23}}{b}_{{2}}{b}_{{3}%
}-{a}_{{22}}{b}_{{3}}^{2}\right) s^{2}+{x\lambda }_{1}^{I}\right) ^{2}$%
\newline

$\beta =\left( \left( -3{a}_{{33}}{b}_{{2}}^{2}+6{a}_{{23}}{b}_{{2}}{b}_{{3}%
}-3{a}_{{22}}{b}_{{3}}^{2}\right) s^{2}+{x\lambda }_{2}^{I}\right) $

$\gamma =\left( \left( -{a}_{{33}}{b}_{{2}}^{2}+2{a}_{{23}}{b}_{{2}}{b}_{{3}%
}-{a}_{{22}}{b}_{{3}}^{2}\right) s^{2}+x\lambda _{3}^{I}\right) $\newline

Therefore $\lambda _{1}^{I}=\lambda _{3}^{I}$ (the root of multiplicity
3)~and the characteristic equation is as follows:

$\left( \left( -{a}_{{33}}{b}_{{2}}^{2}+2{a}_{{23}}{b}_{{2}}{b}_{{3}}-{a}_{{%
22}}{b}_{{3}}^{2}\right) s^{2}+{x\lambda }_{1}^{I}\right) ^{3}\left( 3\left(
-{a}_{{33}}{b}_{{2}}^{2}+2{a}_{{23}}{b}_{{2}}{b}_{{3}}-{a}_{{22}}{b}_{{3}%
}^{2}\right) s^{2}+{x\lambda }_{2}^{I}\right) =0\cdot $

It leads to: $\lambda _{1}^{I}=$ $\frac{\left( a{_{{33}}b}_{{2}}^{2}-2{a}_{{%
23}}{b}_{{2}}{b}_{{3}}+{a}_{{22}}{b}_{{3}}^{2}\right) s^{2}}{x}$ as the
triple multiplicity root and $\lambda _{2}^{I}=\frac{3\left( a{_{{33}}b}_{{2}%
}^{2}-2{a}_{{23}}{b}_{{2}}{b}_{{3}}+{a}_{{22}}{b}_{{3}}^{2}\right) s^{2}}{x}%
=3\lambda _{1}^{I}$ as a single multiplicity root for $s\neq 0$ .

The additional condition $-4{a}_{{23}}^{2}c+4{a}_{{22}}{a}_{{33}}c=0$ we
shall call "polynomial constraint" which relates the coefficients of the
metric in the following way:
\begin{equation}
a_{23}^{2}=a_{22}a_{33}
\end{equation}

This results in $\lambda _{2}^{I}=3\lambda _{1}^{I}$ and therefore
barotropic factor $w=-1/3$ (Quantum Universe with spatial curvature type
barotropic factor):

\begin{equation}
\left( G^{ab}\right) _{_{I}}=\left(
\begin{array}{cccc}
-\lambda ^{I} & 0 & 0 & 0 \\
0 & \frac{1}{3}\lambda ^{I} & 0 & 0 \\
0 & 0 & \frac{1}{3}\lambda ^{I} & 0 \\
0 & 0 & 0 & \frac{1}{3}\lambda ^{I}%
\end{array}%
\right)
\end{equation}

And the corresponding metric tensor:

\begin{equation}
g_{I}=\left(
\begin{array}{cccc}
r^{2s}c & r^{\left( s-1\right) }b_{1} & r^{s}b_{2} & r^{s}b_{3} \\
r^{\left( s-1\right) }b_{1} & \frac{a_{11}}{r^{2}} & \frac{a_{12}}{r} &
\frac{a_{13}}{r} \\
r^{s}b_{2} & \frac{a_{12}}{r} & a_{22} & \sqrt{a_{22}a_{33}} \\
r^{s}b_{3} & \frac{a_{13}}{r} & \sqrt{a_{22}a_{33}} & a_{33}%
\end{array}%
\right)
\end{equation}

II. Another solution is obtained with the polynomial constraint of the form:
$-4{a}_{{23}}^{2}c+4{a}_{22}{a}_{{33}}c=2\left( {a}_{{33}}{b}_{{2}}^{2}-2{a}%
_{{23}}{b}_{{2}}{b}_{{3}}+{a}_{{22}}{b}_{{3}}^{2}\right) $

With this we get the following:

$\alpha =\left( \left( -{a}_{{33}}{b}_{{2}}^{2}+2{a}_{{23}}{b}_{{2}}{b}_{{3}%
}-{a}_{{22}}{b}_{{3}}^{2}\right) s^{2}+{x\lambda }_{1}\right) ^{2}$\newline
$\beta =\left( \left( -{a}_{{33}}{b}_{{2}}^{2}+2{a}_{{23}}{b}_{{2}}{b}_{{3}}-%
{a}_{{22}}{b}_{{3}}^{2}\right) s^{2}+{x\lambda }_{2}\right) $\newline
$\gamma =\left( \left( {a}_{{33}}{b}_{{2}}^{2}-2{a}_{{23}}{b}_{{2}}{b}_{{3}}+%
{a}_{{22}}{b}_{{3}}^{2}\right) s^{2}+x\lambda _{3}\right) $\newline

which leads to:

- the root of multiplicity 3 is $\lambda _{1}^{II}=\frac{\left( {a}_{{33}}{b}%
_{{2}}^{2}-2{a}_{{23}}{b}_{{2}}{b}_{{3}}+{a}_{{22}}{b}_{{3}}^{2}\right) s^{2}%
}{x}=\lambda _{2}^{II}$

- the root of multiplicity 1 is $\lambda _{3}^{II}=-\frac{\left( {a}_{{33}}{b%
}_{{2}}^{2}-2{a}_{{23}}{b}_{{2}}{b}_{{3}}+{a}_{{22}}{b}_{{3}}^{2}\right)
s^{2}}{x}=-\lambda _{1}^{II}$

and contributes to a barotropic factor $w=-1$ (Universe with dark energy -
cosmological constant) with the following Einstein tensor:%
\begin{equation}
\left( G^{ab}\right) _{II}=\left(
\begin{array}{cccc}
-\lambda ^{II} & 0 & 0 & 0 \\
0 & \lambda ^{II} & 0 & 0 \\
0 & 0 & \lambda ^{II} & 0 \\
0 & 0 & 0 & \lambda ^{II}%
\end{array}%
\right)
\end{equation}

From the additional condition $-4{a}_{{23}}^{2}c+4{a}_{22}{a}_{{33}%
}c=2\left( {a}_{{33}}{b}_{{2}}^{2}-2{a}_{{23}}{b}_{{2}}{b}_{{3}}+{a}_{{22}}{b%
}_{{3}}^{2}\right) $ we get the relation for the coefficients of the metric
in the following way:
\begin{equation}
c=-\frac{\left( {a}_{{33}}{b}_{{2}}^{2}-2{a}_{{23}}{b}_{{2}}{b}_{{3}}+{a}_{{%
22}}{b}_{{3}}^{2}\right) }{2\left( {a}_{{23}}^{2}-{a}_{22}{a}_{{33}}\right) }
\end{equation}

Therefore the metric corresponding to this case is:

\begin{equation}
g_{II}=\left(
\begin{array}{cccc}
-r^{2s}\frac{\left( {a}_{{33}}{b}_{{2}}^{2}-2{a}_{{23}}{b}_{{2}}{b}_{{3}}+{a}%
_{{22}}{b}_{{3}}^{2}\right) }{2\left( {a}_{{23}}^{2}-{a}_{22}{a}_{{33}%
}\right) } & r^{\left( s-1\right) }b_{1} & r^{s}b_{2} & r^{s}b_{3} \\
r^{\left( s-1\right) }b_{1} & \frac{a_{11}}{r^{2}} & \frac{a_{12}}{r} &
\frac{a_{13}}{r} \\
r^{s}b_{2} & \frac{a_{12}}{r} & a_{22} & a_{23} \\
r^{s}b_{3} & \frac{a_{13}}{r} & a_{23} & a_{33}%
\end{array}%
\right)
\end{equation}

\section*{Acknowledgments}

The work by T.J. and S.M. has been fully supported by Croatian Science
Foundation under the project (IP-2014-09-9582). T.J. would like to thank B.
Klajn for the fruitful discussions that led to Appendix 1. A. Much is
acknowledged for the discussions with T.J. A.P. acknowledges support from
the European Union's Seventh Framework programme for research and innovation
under the Marie Sk\l odowska-Curie grant agreement No 609402 - 2020
researchers: Train to Move (T2M). Part of this work was also supported by
Polish National Science Center project 2014/13/B/ST2/04043. AB is supported
by NCN project DEC-2013/09/B/ST2/03455.


\begin{thebibliography}{99}
\bibitem{DFR94} S.~Doplicher, K.~Fredenhagen and J.~E.~Roberts,
\textquotedblleft Space-time quantization induced by classical
gravity,\textquotedblright\ Phys.\ Lett.\ B 331, 39 (1994).

\bibitem{DFR95} S.~Doplicher, K.~Fredenhagen and J.~E.~Roberts,
\textquotedblleft The Quantum structure of space-time at the Planck scale
and quantum fields,\textquotedblright\ Commun.\ Math.\ Phys.\ 172, 187
(1995), [hep-th/0303037].

\bibitem{2} S.~Majid and H.~Ruegg, \textquotedblleft Bicrossproduct
structure of kappa Poincar\'e group and noncommutative
geometry,\textquotedblright\ Phys.\ Lett.\ B 334, 348 (1994),
[hep-th/9405107].

\bibitem{Zakrzewski} S. Zakrzewski, \textquotedblleft Quantum Poincar\'e group
related to the kappa -Poincar\'e algebra\textquotedblright\ J. Phys. A 27,
2075 (1994)

\bibitem{Schupp} P.~Schupp and S.~Solodukhin, \textquotedblleft Exact Black
Hole Solutions in Noncommutative Gravity,\textquotedblright\ arXiv:0906.2724
[hep-th].

\bibitem{Schenkel} T.~Ohl and A.~Schenkel, \textquotedblleft Cosmological
and Black Hole Spacetimes in Twisted Noncommutative
Gravity,\textquotedblright\ JHEP 0910, 052 (2009), [arXiv:0906.2730
[hep-th]]. \newline
T.~Ohl and A.~Schenkel, \textquotedblleft Symmetry Reduction in Twisted
Noncommutative Gravity with Applications to Cosmology and Black
Holes,\textquotedblright\ JHEP 0901, 084 (2009), [arXiv:0810.4885 [hep-th]].
\newline
A.~Schenkel, PhD Thesis \textquotedblleft Noncommutative Gravity and Quantum
Field Theory on Noncommutative Curved Spacetimes,\textquotedblright\
arXiv:1210.1115 [math-ph].

\bibitem{Mairi} W.~Nelson and M.~Sakellariadou, \textquotedblleft Cosmology
and the Noncommutative approach to the Standard Model,\textquotedblright\
Phys.\ Rev.\ D 81, 085038 (2010), [arXiv:0812.1657 [hep-th]].

\bibitem{BTZ} K.~S.~Gupta, S.~Meljanac and A.~Samsarov, \textquotedblleft
Quantum statistics and noncommutative black holes,\textquotedblright\ Phys.\
Rev.\ D 85, 045029 (2012), [arXiv:1108.0341 [hep-th]]. \newline
K.~S.~Gupta, E.~Harikumar, T.~Juri\'{c}, S.~Meljanac and A.~Samsarov,
\textquotedblleft Noncommutative scalar quasinormal modes and quantization
of entropy of a BTZ black hole,\textquotedblright\ JHEP 1509, 025 (2015),
[arXiv:1505.04068 [hep-th]]. \newline
K.~S.~Gupta, E.~Harikumar, T.~Juric, S.~Meljanac and A.~Samsarov,
\textquotedblleft Effects of Noncommutativity on the Black Hole
Entropy,\textquotedblright\ Adv.\ High Energy Phys.\ 2014, 139172 (2014),
[arXiv:1312.5100 [hep-th]].

\bibitem{link} %
http://www.ast.cam.ac.uk/research/cosmology.and.fundamental.physics/gravitational.waves

\bibitem{1} J. Lukierski, A. Nowicki, H. Ruegg, V. N. Tolstoy, ''Q deformation
of Poincar\'e algebra'', Phys. Lett. B 264 (1991) 331;\newline
J. Lukierski, H. Ruegg, ''Quantum $\kappa $-Poincar\'e in Any Dimensions'', Phys.
Lett. B 329 (1994) 189

\bibitem{bgmp10} A. Borowiec, K. S. Gupta, S. Meljanac and A. Pacho\l ,
\textquotedblleft Constraints on the quantum gravity scale from
kappa--Minkowski spacetime\textquotedblright , Europhys. Lett. 92, 20006
(2010), arXiv:0912.3299.

\bibitem{hajume} E.~Harikumar, T.~Juric and S.~Meljanac, ''Geodesic equation
in $\kappa $-Minkowski spacetime,\textquotedblright\ , Phys. Rev. D 86,
045002 (2012) arXiv:1203.1564 [hep-th].

\bibitem{Aschieri} P. Aschieri, C. Blohmann, M. Dimitrijevi\'{c} , F. Meyer,
P. Schupp, J. Wess, ''A gravity theory on noncommutative spaces'', Classical Quantum Gravity 22, 3511 (2005).\newline
P. Aschieri, M. Dimitrijevi\'{c}, F. Meyer, J. Wess, ''Noncommutative Geometry and Gravity'', Classical Quantum
Gravity 23, 1883 (2006).\newline
P.~Aschieri, \textquotedblleft Noncommutative Gravity and the *-Lie algebra
of diffeomorphisms,\textquotedblright\ Subnucl.\ Ser.\ 44, 519 (2008)
[hep-th/0703014].

\bibitem{Woronowicz1} P. Podles, L. Woronowicz, ''On the classification of quantum Poicar\'{e} groups'', Commun. Math. Phys. 178 (1996),
[arXiv:hep-th/9412059].

\bibitem{Sitarz} A.~Sitarz, \textquotedblleft Noncommutative differential
calculus on the kappa Minkowski space,\textquotedblright\ Phys.\ Lett.\ B
349, 42 (1995) [hep-th/9409014]

\bibitem{Gonera} C. Gonera, P. Kosinski and P. Maslanka, ''Differential
calculi on quantum Minkowski space'', J. Math. Phys. 37,5820 (1996)

\bibitem{Mercati} F.~Mercati,
  ``Quantum $\kappa$-deformed differential geometry and field theory'',
  Int.\ J.\ Mod.\ Phys.\ D {\bf 25}, no. 05, 1650053 (2016)
  [arXiv:1112.2426 [math.QA]].

\bibitem{hep-th/0307038} K. Przanowski, \textquotedblleft The bicovariant
differential calculus on the kappa-Poincar\'e and kappa-Weyl
groups\textquotedblright\ Czech Journ. Phys. 47, 107, (1997), q-alg/9606022%
\newline
P. Podles, \textquotedblleft Solutions of Klein--Gordon and Dirac equations
on quantum Minkowski spaces\textquotedblright , Commun. Math. Phys. 181,
569-586 (1996), q-alg/9510019\newline
P. Kosinski, P. Maslanka, J. Lukierski, A. Sitarz, ``Generalized kappa-Deformations and Deformed Relativistic Scalar Fields on
Noncommutative Minkowski Space'', Proceedings of the
Conference "Topics in Mathematical Physics, General Relativity and
Cosmology", World Scientific, 2003, arXiv:hep-th/0307038

\bibitem{toward} T.~Juric, S.~Meljanac, D.~Pikutic and R.~Strajn, ``Toward
the classification of differential calculi on $\kappa$-Minkowski space and
related field theories,'' JHEP 1507, 055 (2015), arXiv:1502.02972
[hep-th].

\bibitem{Bu} J.~G.~Bu, J.~H.~Yee and H.~C.~Kim, \textquotedblleft
Differential Structure on kappa-Minkowski Spacetime Realized as Module of
Twisted Weyl Algebra,\textquotedblright\ Phys.\ Lett.\ B 679, 486 (2009),
[arXiv:0903.0040 [hep-th]].

\bibitem{KJ} S.~Meljanac and S.~Kresic-Juric, \textquotedblleft
Noncommutative Differential Forms on the kappa-deformed
Space,\textquotedblright\ J.\ Phys.\ A 42, 365204 (2009), [arXiv:0812.4571
[hep-th]]. \newline
S.~Meljanac and S.~Kresic-Juric, \textquotedblleft Differential structure on
kappa-Minkowski space, and kappa-Poincar\'e algebra,\textquotedblright\ Int.\
J.\ Mod.\ Phys.\ A 26, 3385 (2011), [arXiv:1004.4647 [math-ph]].

\bibitem{41} S. Meljanac, S. Kresic-Juric ,R. Strajn, ''Differential algebras
on kappa-Minkowski space and action of the Lorentz algebra'', Int. J. Mod.
Phys. A27 (2012) 1250057 ;

\bibitem{EPJC} T. Juric, S. Meljanac, R. \v{S}trajn, ''Differential forms and
k-Minkowski spacetime from extended twist\textquotedblright , Eur. Phys. J.
C (2013) 73: 2472, arXiv:1211.6612 [hep-th]

\bibitem{oeckl} R. Oeckl, ''Classification of Differential Calculi on $%
U_{q}(b_{+})$, Classical Limits, and Duality'', J. Math. Phys. 40, 3588-3604,
1999, arXiv:math/9807097 [math.QA]

\bibitem{majidbeggs} E.~J.~Beggs and S.~Majid, \textquotedblleft Gravity
induced from quantum spacetime,\textquotedblright\ Class.\ Quant.\ Grav.\
31, 035020 (2014), [arXiv:1305.2403 [gr-qc]].

\bibitem{majid2014} S.~Majid and W.~Q.~Tao, \textquotedblleft Cosmological
constant from quantum spacetime\textquotedblright\, Phys. Rev. D91:124028, 2015, arXiv:1412.2285 [gr-qc].

\bibitem{majid2} S.~Majid and W.~Q.~Tao, \textquotedblleft Noncommutative
Differentials on Poisson-Lie groups and pre-Lie algebras,\textquotedblright\
Pacific Journal of Mathematics -- in press, arXiv:1412.2284 [math.QA].

\bibitem{BP2014} A. Borowiec, A. Pachol, \textquotedblright $\kappa $%
-Deformations and Extended $\kappa $-Minkowski Spacetimes\textquotedblright
, SIGMA 10 (2014), 107, arXiv:1404.2916

\bibitem{BP2009} A. Borowiec, A. Pachol, \textquotedblright $\kappa -$%
Minkowski spacetime as the result of Jordanian twist
deformation\textquotedblright , Phys.Rev.D79:045012,2009, arXiv:0812.0576

\bibitem{Vitale} E. Di Grezia, G. Esposito, P. Vitale \textquotedblright
Self-dual road to noncommutative gravity with twist: a new
analysis\textquotedblright , Phys.Rev. D89 (2014) 6, 064039, Phys.Rev. D90
(2014) 12, 129901, arXiv:1312.1279
\end{thebibliography}
\end{document}